\documentstyle[psfig,12pt,aaspp4]{article}
\def\Ntot{90}
\def\Nopt{82}
\def\kms{km~s$^{-1}$}
\def\arcmin{$^{\prime}$}
\def\arcsec{$^{\prime\prime}$}
\def\hal{H$\alpha$}
\def\be{\begin{equation}}
\def\ee{\end{equation}}
\def\about{$\sim$}
\def\etal{{\it et al.}}
\def\eg{e.g.}
\def\ie{i.e.}

\def\rA{$R_{\rm A}$}
\def\Ropt{$R_{\rm opt}$}
\def\Wobs{$W_{\rm obs}$}

\def\HI{\ion{H}{1}}
\def\HII{\ion{H}{2}}
\def\NII{[\ion{N}{2}]}

\begin{document}
\hskip 3.5in{\hskip 10pt \date{5 November 1997}}
\title{SEEKING THE LOCAL CONVERGENCE DEPTH. II. TULLY--FISHER OBSERVATIONS OF THE 
CLUSTERS A114, A119, A194, A2295, A2457, A2806, A3193, A3381, AND A3744}
 
\author {DANIEL A. DALE,\altaffilmark{1,2} RICCARDO GIOVANELLI,\altaffilmark{1,2} 
MARTHA P. HAYNES,\altaffilmark{1} AND MARCO SCODEGGIO\altaffilmark{1,3}}
\affil{Center for Radiophysics and Space Research
and National Astronomy and Ionosphere Center, Cornell University, Ithaca, NY 
14853}
\affil{Electronic mail: dale,riccardo,haynes,scodeggi@astrosun.tn.cornell.edu}

\author {EDUARDO HARDY\altaffilmark{2}}
\affil{Department of Physics, Laval University, Ste--Foy, P.Q., G1K 7P4, Canada}
\affil{Electronic mail: hardy@phy.ulaval.ca}

\author {LUIS E. CAMPUSANO\altaffilmark{2}}
\affil{Observatorio Astron\'{o}mico Cerro Cal\'{a}n, Departamento de 
Astronom\'{\i}a, Universidad de Chile, Casilla 36-D, Santiago, Chile}
\affil{Electronic mail: lcampusa@das.uchile.cl}

\altaffiltext{1}{Visiting Astronomer, Kitt Peak National Observatory, National
Optical Astronomy Observatories, which are operated by the Association of 
Universities for Research in Astronomy, Inc., under a cooperative agreement 
with the National Science Foundation.}
\altaffiltext{2}{Visiting Astronomer, Cerro Tololo Inter--American Observatory,
NOAO.}
\altaffiltext{3}{Now at European Southern Observatory, Karl Schwarzschild Str. 
2, D--85748 Garching b. M\"unchen, Germany.}

\begin{abstract}

We present Tully--Fisher (TF) observations for nine rich Abell clusters of
galaxies.  This is the second such data installment of an all--sky survey of 
\about\ 50 clusters in the redshift range $0.02 \lesssim z \lesssim 0.06$.  The
data extends the TF study of nearby clusters of Giovanelli \etal\ (1997a,b); they
will be used jointly to determine an accurate $I$ band TF template and to
establish a cluster inertial reference frame to $z$ \about\ 0.06.
\end{abstract}


\section {INTRODUCTION}

The accurate measurement of deviations from smooth Hubble flow remains a challenge
in observational cosmology.  The lack of a proven redshift--independent distance 
estimator applicable beyond \about\ 100$h^{-1}$ Mpc becomes sorely obvious when 
claims of large amplitude coherent flows on scales larger than this are made 
(Lauer \& Postman 1992:LP; Courteau \etal\ 1993).  Recent observational efforts at 
constructing relations useful for peculiar velocity measurements (\eg\ Riess 
\etal\ 1995 and Giovanelli \etal\ 1996: G96) challenge the LP result, though with 
insufficient certainty.  The Tully--Fisher (TF) sample of G96 peculiar velocities 
is based on a template defined by 555 galaxies in 24 clusters (Giovanelli \etal\
1997a,b:G97a,b) that reaches only to $cz$ \about\ 9000 \kms\ and is thus unable to 
probe the entire peculiar velocity field the LP flow encompasses, whereas the 
Riess \etal\ sample of supernova type Ia is too sparsely populated to adequately 
characterize the local bulk flow (Watkins \& Feldman 1995).  The existing samples 
of peculiar velocities at relatively large distances ($>$ 100$h^{-1}$) need to be 
enlarged.

We are expanding upon the data of G97a,b by obtaining TF measurements for an
all--sky survey of some 50 clusters in the redshift range 5000 $\lesssim cz 
\lesssim$ 20,000 \kms.  The benefits of such an enterprise are twofold.  First,
the combination of our data set with that of G97a,b will yield a highly accurate 
TF template.  The proximity of the extant G97a,b sample allows a wide range of 
galactic properties to be observed and is thus ideal for determining the slope of 
the TF relation.  On the other hand, the redshifts of our sample of relatively 
distant clusters are less affected by the vagaries of cosmic peculiar motions, a 
property useful for accurately extracting the TF magnitude offset.  Second, the 
combined data set will cover a much larger volume than that of G97a,b. Recent 
work by Scaramella \etal\ (1994), Tini-Brunozzi \etal\ (1995), and Branchini \etal\
(1996) suggests that coherent peculiar motions may persist as far as 18,000 
\kms, \ie\ the local ``convergence depth'' may extend to \about\ 180$h^{-1}$ Mpc.  
It thus appears crucial to penetrate the local velocity field as deeply as 
possible:  our sample extends out to and possibly beyond the local convergence 
depth.  This will enable dipole motion measurements over scales large enough to 
test the LP claim, whose sample has an effective depth of \about\ 100$h^{-1}$.

In order to provide public access to the data on the shortest possible time 
scale, we are presenting results of our survey in installments, as we progress in 
the data reductions for sizable fractions of the cluster set.  Dale \etal\ (1997;
hereafter ``Paper I'') presented TF measurements for 84 galaxies in the fields of
seven Abell clusters.  In this paper we present the project's second data 
installment: TF measurements for \Ntot\ galaxies in the fields of nine Abell 
clusters, one of which, the field of A2295, contains two separate clusters (see 
Sec. 3).  The following section reviews the imaging and spectroscopic observations 
for this data installment.  Sec. 3 presents the relevant TF data. 

\section {OBSERVATIONS}					% section 2

Imaging for this project began in October 1994 and is now complete, whereas 
spectroscopy measurements began in December 1995 and are still ongoing.  In the 50
clusters we have chosen to include in our all--sky survey, we plan to obtain 5 to
15 TF measurements per cluster.  Including approximately two dozen 21 cm line 
widths taken earlier by RG and MH, we have thus far obtained \about\ 360 (cluster
member) velocity widths in 48 Abell clusters; we expect to measure \about\ 100 more
velocity widths.  Paper III of this series will present the final installment of
data for the remaining clusters.

\subsection {$I$ Band Imaging}

All photometric observations were carried out in the $I$ band (Kron--Cousins 
filter, central wavelength of 8075 \AA\ and passband of 1500 \AA), matching those 
in Paper I and G97a,b.  The imaging for the clusters in this paper was carried out 
at the KPNO and CTIO 0.9 m telescopes.  Exposures for each frame amounted to 600 
seconds, a time sufficient to reliably reach isophotal levels near 24.5 mag 
arcsec$^{-2}$.  The clusters A119, A194, A2295, and A2457 were observed at KPNO 
over the course of several observing runs:  October 5--16 1994, September 14--21 
1995, September 21--October 1 1995, November 20--26 1995, October 31--November 12 
1996, and February 4--19 1997.  The clusters A2806, A114, A3193, A3381, and A3744 
were observed at CTIO during the runs of February 1--8 1995 and 
August 28--September 4 1995.  We used the same observational methods and technical 
setups described in Paper I for the KPNO observations; the only significant 
difference between imaging at KPNO and CTIO for our cluster sample is the higher 
spatial resolution for the data taken at the CTIO 0.9 m (0.4\arcsec\ pixel$^{-1}$ 
versus 0.68\arcsec\ pixel$^{-1}$), and a smaller field of view (14\arcmin\ $\times$
14\arcmin\ versus 23\arcmin\ $\times$ 23\arcmin) for the 2048$^2$ CCD.  The average
seeing FWHM for the images used in this data set was 1.7\arcsec\ $\pm$ 0.2\arcsec\ 
at KPNO and 1.4\arcsec\ $\pm$ 0.2\arcsec\ at CTIO; however, the nights with the 
best seeing conditions were preferentially devoted to the more distant clusters.  
The majority of the data presented here were taken in good photometric conditions, 
for which the photometric zero point calibration could be determined with an 
accuracy of 0.02 mag or better.  In a minority of cases (2\%), photometric 
conditions were of inferior quality.  In those cases frames were taken with a 
substantial sky overlap with images taken in good photometric conditions, so that 
fluxes of at least 12 field stars could be measured in the overlap region, thus 
guaranteeing calibration to the \about\ 0.03 mag level.

The determination of $I$ band fluxes follows from data reduction methods discussed
in Paper I and Haynes \etal, in preparation, using both standard and customized
IRAF\footnote{IRAF (Image Reduction and Analysis Facility) is distributed by NOAO.} packages.  We will only mention here that
the measured fluxes, denoted $m_{\infty}$, include extrapolations of the 
exponential fits to the surface brightness profiles to infinity and are typically accurate to \about\ 0.03 mag (uncertainties at least this large are later included 
after corrections for internal extinction are made; see Paper I for details on flux
errors).  We apply several corrections to $m_{\infty}$, to obtain:
\be
m_{\rm cor} = m_{\infty} - A_I + k_I - \Delta m_{\rm int} - \Delta m_T.
\ee
The first correction, $A_I$, is for the extinction caused by the Milky Way.  We
use Burstein and Heiles' (1978) tabulation of galactic extinction values by 
averaging entries in the {\it Third Reference Catalogue of Bright Galaxies} (de 
Vaucouleurs \etal\ 1991) found near the cluster centers.  We convert $B$ band 
results to $I$ band via $A_I$ = 0.45$A_B$; values of $A_I$ for the nine clusters 
presented here range from 0.00 to 0.17 mag.  The internal extinction correction, 
$\Delta m_{\rm int}$, is applied using the procedure outlined in G97a,
\be
 \Delta m_{\rm int} = - f(T) \; \gamma(W_{\rm cor}) \; \log(1-e),
\ee
where $\gamma$ ($\lesssim$ 1.0) depends on the corrected velocity width 
$W_{\rm cor}$ (Sec. 2.2) and $e$ is the ellipticity of the spiral disk, corrected 
for atmospheric seeing effects as described in Sec. 5 of Paper I (the adopted 
correction $\Delta m_{\rm int}$ is slightly smaller for early, less dusty galaxies:
$f(T)$=0.85 for types $T$ earlier than Sbc; $f(T)$=1 otherwise).  We apply a 
cosmological k-correction: $k_I = (0.5876 - 0.1658T)z$ (Han 1992).  Finally, we 
include a small correction, $\Delta m_T$, for the TF dependence on morphological 
type found in G97b; while the true, unextincted, face--on apparent magnitude of the
galaxy is $m_{\rm cor} + \Delta m_T, m_{\rm cor}$ represents the value to be used 
in TF work with the template TF relation obtained in G97b, characterized for Sbc 
and Sc galaxies.  Thus $\Delta m_T = 0$ for types Sbc and later; the correction is 
0.1 mag for Sb types and 0.32 mag for types earlier than Sb.

\subsection {Optical Spectroscopy}

Rotational velocity widths for this sample of cluster galaxies were extracted from
long--slit spectra obtained at the Palomar Observatory 5 m telescope and the CTIO 
4 m telescope.  The clusters A114, A119, A194, A2295, A3744, and A2457 were 
observed at Palomar during the runs of October 22--23 1995, December 13--18 1995, 
July 9--12 1996, and September 13--19 1996.  The clusters A2806, A3193, and A3381 
were observed at CTIO during the nights of April 4--7 1996 and 
September 30--October 3 1996.  We refer the reader to Paper I for details of the 
spectroscopy performed at Palomar.  The observing setup at the CTIO 4 m uses the 
R--C Spectrograph and the Loral 3k CCD with a 203\arcsec\ long slit.  The 
combination of the 1200 lines mm$^{-1}$ grating and a 2\arcsec\ wide slit yields a 
dispersion of 0.55 \AA\ pixel$^{-1}$ and a spectral resolution of 1.7 \AA\ 
(equivalent to 75 \kms\ at 6800 \AA).  Along the cross--dispersion axis, the 
spatial scale is 0.50\arcsec\ pixel$^{-1}$.  The wavelength range (\about\ 5640 
\AA\ to 7320 \AA) is large enough to observe redshifted \hal\ emission in galaxies 
with recessional velocities up to \about\ 34,000 \kms, in our chosen setup.  

The adopted observing strategy includes a five minute preliminary integration on each galaxy.  This allows a quick estimate of the spectral lines' strength from 
which we determine the exposure time necessary to adequately sample the outer disk 
regions.  Moreover, performing a five minute test exposure informs us if the
galaxy is at the cluster redshift and if the distribution of \HII\ regions is
sufficient to provide a useful rotation curve.  If the observation is deemed 
useful, a second integration of usually 15--45 minutes is taken. 

We extract optical rotation curves (ORCs) as discussed in Paper I.  We use the 
\hal\ (6563 \AA) emission line in mapping out the ORC except in 2\% of the cases 
where the emission of the \NII\ (6584 \AA) line extends to larger 
radii than that of the \hal\ emission.  We kinematically center the ORC by 
assigning the velocity nearest the average of the 10\% and 90\% velocities to be 
at radius zero, where an N\% velocity is such that N\% of the data points in the
rotation curve have velocity smaller than it.  Besides representing the velocity at
the spatial center of the galaxy, we also assume the velocity of the ORC at this 
radius to be the galaxy's recessional velocity.  We define the observed rotational 
velocity width to be \Wobs\ $\equiv V_{\rm 90\%} - V_{\rm 10\%}$.  In regions where
heavy \hal\ absorption and emission are mixed, usually near the galactic center, we
fill in the portions of the rotation curve with data from the \NII\ rotation curve, if available.  
We do this to provide information on the shape of the inner portions of all ORCs 
and to ensure that our method of computing \Wobs\ does not artificially yield 
slightly large values when data is missing.  We apply this \NII\ patch to 
approximately 10\% of the \hal\ ORCs; the notes on individual galaxies in Sec. 3 
indicates which ORCs include \NII\ patches.  In Fig. 1 we display the effect a 
\NII\ patch can have on a ORC using the ORC of galaxy 400641 in Abell 119 as an
example.  The upper panel gives the \hal\ ORC without the \NII\ patch, the lower 
panel includes the patch.  The effect on the inferred redshift and \Wobs\ is small,
but the shape of the ORC has changed significantly.  This change is in part due to the kinematical recentering of the ORC --- the origin of the ORC has also been 
shifted radially (by 0.94\arcsec).  If information on the central portions of an 
ORC is lacking for all spectral lines available, we consider the peak of the 
galactic continuum light profile, formed by summing the data along the dispersion 
direction, to represent the spatial center of the galaxy (see, for example, the
upper panel of Fig. 1).

The method described above on characterizing rotational velocity widths is used
since it effectively ignores small--scale velocity irregularities within an ORC 
that may arise from streaming motions within the galactic disk, the non--uniform 
distribution of \HII\ regions, or distortions associated with the spiral 
pattern.  Furthermore, we find empirically that provided that the rotation curve 
extends far enough out into the disk, this method recovers the velocity width at 
\Ropt, the distance along the major axis to the isophote containing 83\% of the $I$
band flux.  This radius is reported by Persic \& Salucci (1991 \& 1995) to be a 
useful radius at which to measure the velocity width of ORCs.  Thus, extrapolations
to the ORC, and hence adjustments to \Wobs, are made in cases where the ORC's 
radial extent falls well short of \Ropt.  The resulting correction which depends on
the shape of the rotation curve, ${\Delta}_{\rm sh}$, rarely exceeds 0.1\Wobs.

To recover the actual velocity widths, two more corrections are necessary.  The 
first is the factor 1/sin$i$ necessary to convert the width observed for a disk 
inclined to the line of sight at an angle $i$ to what would be observed if the disk
were edge--on, and the second is the factor 1/(1+$z$) to correct for the 
cosmological broadening of $W$.  The corrected optical rotational velocity width is
then
\be
W_{\rm cor} = {{W_{\rm obs} + {\Delta}_{\rm sh}} \over {(1+z)\sin i}}.
\ee
A discussion of the errors in the velocity widths can be found in Paper I.   

In addition to the \Nopt\ ORCs presented here, we also have 21 cm measurements for
eight galaxies, all members of Abell 194.   Six of the eight \HI\ spectra were
taken at the Arecibo Observatory using a spectrometer channel separation of 8 \kms\
while the remaining two were obtained at the Greenbank 300' telescope (Haynes \&
Giovanelli 1991) at a spectral resolution of 11 \kms.  A typical signal to noise 
ratio for these observations was 10; errors in the observed velocity widths are of 
order 15 \kms.  Details on the \HI\ data reduction and the corrections made to the 
observed velocity widths can be found in G97a.  A full discussion of the comparison
between \HI\ and optical widths, is given in Giovanelli \etal\ (1997c).

\section {DATA}					% section 6

Table 1 lists the main parameters of the clusters.  Standard names are listed in 
column 1.  Adopted coordinates of the cluster center are listed in columns 2 and 
3, for the epoch 1950; they are obtained from Abell \etal\ (1989), except for the entry A2295b, a system found to be slightly offset from A2295 in both sky position 
and redshift.  For all the clusters we derived a new systemic velocity, combining 
the redshift measurements available in the NED\footnote{The NASA/IPAC Extragalactic
Database is operated by the Jet Propulsion Laboratory, California Institute of 
Technology, under contract with the National Aeronautics and Space Administration.}
database with our own measurements.  These newly determined velocities are listed 
in columns 4 and 5, in the heliocentric and in the cosmic microwave background 
(CMB) reference frame (Kogut \etal\ 1993), respectively.  We list the number of 
cluster member redshifts used
in determining systemic velocities in column 6.  An estimated error for the 
systemic velocity is parenthesized after the heliocentric figure.  Spherical and 
Cartesian supergalactic coordinates are given in columns 7 and 8, and in columns 
9--11, respectively.

Figs. 2--6 show the distribution of the galaxies in each cluster.  The top panel in
each of these figures displays the spatial location of:  the outline of the fields 
imaged (large squares), cluster members (circles --- those with poor/unusable 
velocity widths are left unfilled), background or foreground objects (asterisks), 
and galaxies with known redshift but without reliable widths (dots).  Circles of 1 
and 2 Abell radii, \rA, are drawn as dashed lines, if the area displayed is large 
enough.  If no dashed circle is drawn, \rA\ is larger than the figure limits.  We 
also plot radial (CMB) velocity as a function of angular distance from the cluster 
center in the lower panel of each figure.  A dashed vertical line is drawn at 1 
\rA.  The combination of the sky and velocity plots is used to gauge cluster 
membership for each galaxy.  

We separate photometric data and spectroscopic data in two tables.  Table 2 lists 
the spectroscopic properties and Table 3 gives the pertinent photometric results.  

Entries in the tables are sorted first by the Right Ascension of each cluster, 
and within each cluster sample by increasing galaxy Right Ascension.  The listed 
parameters for Table 2 are:

\noindent 
Col 1: identification names corresponding to a coding number in our private
database, referred to as the Arecibo General Catalog, which we maintain for easy 
reference in case readers may request additional information on the object.

\noindent
Cols. 2 and 3: Right Ascension and Declination in the 1950.0 epoch.  Coordinates
have been obtained from the Digitized Sky Survey catalog and are accurate to
$<$ 2\arcsec.

\noindent
Cols. 4 and 5: the galaxy radial velocity as measured in the heliocentric and
CMB reference frame (Kogut \etal\ 1993).  Errors are parenthesized: \eg\ 
13241(08) implies 13241$\pm$08.

\noindent
Col. 6: the raw velocity width in \kms. Measurement of optical widths are 
described in Sec. 2.2; 21 cm line widths are denoted with a dagger and refer to 
values measured at a level of 50\% of the profile horns.

\noindent
Col. 7: the velocity width in \kms\ after correcting for ORC shape, the 
cosmological stretch of the data and, for 21 cm data, signal to noise effects, 
insterstellar medium turbulence, and instrumental and data processing broadening; 
details on the adopted corrections for optical and 21 cm data are given in Sec. 2.2
and G97a, respectively.

\noindent
Col. 8:  the corrected velocity width in \kms\ converted to an edge--on perspective.

\noindent
Col. 9: the adopted inclination $i$ of the plane of the disk to the line of 
sight, in degrees, (90$^\circ$ corresponding to edge--on perspective); the 
derivation of $i$ and its associated uncertainty are discussed in Sec. 4 of 
Paper I.

\noindent
Col. 10: the logarithm in base 10 of the corrected velocity width (value in 
column 7), together with its estimated uncertainty between brackets. The 
uncertainty takes into account both measurement errors and uncertainties arising 
from the corrections. The format 2.576(22), for example, is equivalent to 
2.576$\pm$0.022.

The position angle adopted for the slit of each spectroscopic observation is 
that given in column 4 of Table 3.  The first column in Table 3 matches that of 
Table 2.  The remaining listed parameters for Table 3 are:

\noindent
Col. 2: morphological type code in the RC3 scheme, where code 1 corresponds 
to Sa's, code 3 to Sb's, code 5 to Sc's and so on.  When the type code is 
followed by a ``B'', the galaxy disk has an identifiable bar.  We assign these
codes after visually inspecting the CCD $I$ band images and after noting the value
of $R_{\rm 75}/R_{\rm 25}$, where $R_X$ is the radius containing X\% of the $I$ 
band flux.  This ratio is a measure of the central concentration of the flux which 
was computed for a variety of bulge--to--disk ratios.  Given the limited 
resolution of the images, some of the inferred types are rather uncertain; 
uncertain types are followed by a colon.

\noindent
Col. 3: the angular distance $\theta$ in arcminutes from the center of each 
cluster.

\noindent
Col. 4: position angle of the major axis of the image, also used for 
spectrograph slit positioning (North: 0$^{\circ}$, East: 90$^{\circ}$).

\noindent
Col. 5: observed ellipticity of the disk.
 
\noindent
Col. 6: ellipticity corrected for seeing effects as described in Sec. 5 of 
Paper I, along with its corresponding uncertainty.  The format 0.780(16), for 
example, is equivalent to 0.780$\pm$0.016.

\noindent
Col. 7: surface brightness of the disk at zero radius, as extrapolated from the fit
to the disk surface brightness profile.

\noindent
Col. 8: the (exponential) disk scale length.

\noindent
Col. 9: the distance along the major axis to the isophote containing 83\% of the 
$I$ band flux.

\noindent
Col. 10: isophotal radius along the major axis where the surface brightness 
equals 23.5 mag sec$^{-2}$.

\noindent
Col. 11: apparent magnitude within the 23.5 mag sec$^{-2}$ isophote.

\noindent
Col. 12: the measured $I$ band magnitude, extrapolated to infinity assuming 
that the surface brightness profile of the disk is well described by an 
exponential function.

\noindent
Col. 13: the apparent magnitude, to which k--term, galactic and internal
extinction corrections were applied; details on the adopted corrections are 
given in Sec. 2.1. 

\noindent
Col. 14: the absolute magnitude, computed assuming that the galaxy is at the 
distance indicated either by the cluster redshift, if the galaxy is a cluster 
member within 1 \rA\ of the cluster center, or by the galaxy redshift if not.  The 
calculation assumes $H_\circ = 100h$ \kms\ Mpc$^{-1}$, so the value listed is 
strictly $M_{\rm cor} - 5\log h$.  In calculating this parameter, radial velocities
are expressed in the CMB frame and uncorrected for any cluster peculiar motion.  
The uncertainty on the magnitude, indicated between brackets in hundredths of a 
mag, is the sum in quadrature of the measurement errors and the estimate of the 
uncertainty in the corrections applied to the measured parameter.

When an asterisk appears at the end of the record, a  detailed comment is given 
for that particular object.  Because of the length and number of these comments, 
they are not appended to the table but included in the text as follows.  Note
that a record is flagged in both tables 2 and 3, independently on whether the
comments refer only to the photometry, only to the spectroscopy, or both.

\noindent {\bf A2806:}\\
\small
\noindent  20391: possible disk warp; \NII\ patch for radii $<$ 8\arcsec\\
\noindent 400698: mostly bulge ORC; uncert. extrap.\\
\noindent 400703: mostly bulge ORC; uncert. extrap.; unfit for TF use.\\
\noindent 400713: mostly bulge ORC; large extrap.; exp. disk over narrow range of 
radii.

\normalsize
\noindent {\bf A114:}\\
\small
\noindent 400672: exp. disk over narrow range of radii.\\
\noindent 400443: uncert. ellipticity; \hal\ absorption in bulge -- \NII\ patch
for radii $<$ 3\arcsec; two nuclei?\\
\noindent 400453: uncert. PA.

\normalsize
\noindent {\bf A119:}\\
\small
\noindent 400619: exp. disk over narrow range of radii; \hal\ absorption in bulge
-- \NII\ patch for radii $<$ 2.5\arcsec.\\
\noindent 400600: uncert. ORC extrap.; \NII\ line
used.\\
\noindent 400641: uncert. ellipticity; \NII\ patch for radii $\lesssim$
2\arcsec.\\
\noindent 400658: background gal.

\normalsize
\noindent {\bf A194:}\\
\small
\noindent 784: only 21 cm width available.\\
\noindent 830: only 21 cm width available; note low i; unfit for TF use.\\
\noindent 894: foreground gal.; only 21 cm width available.\\
\noindent 915: background gal.; only 21 cm width available.\\
\noindent 931: foreground gal.; only 21 cm width available.\\
\noindent 410584: asymm. nucleus.\\
\noindent 1116: only 21 cm width available.\\
\noindent 1120: only 21 cm width available.\\
\noindent 1123: only 21 cm width available.

\normalsize
\noindent {\bf A3193:}\\
\small
\noindent 23113: note low i.\\
\noindent 23130: extremely low surf. brightness at radii $\gtrsim$ 20\arcsec; 
heavy \hal\ absorption throughout disk; unfit for TF use.\\
\noindent 23136: note low i; \NII\ patch for radii $\lesssim$ 0.5\arcsec.

\normalsize
\noindent {\bf A3381:}\\
\small
\noindent 460069: rising ORC; uncert. extrap.\\
\noindent 460071: \hal\ absorption in bulge. \\
\noindent 24895: companion galaxy 35\arcsec\ to SE.

\normalsize
\noindent {\bf A2295b:}\\
\small
\noindent 270370: rising ORC; uncert. extrap.; uncertain PA.\\
\noindent 280128: mostly bulgy ORC.\\
\noindent 280130: star 40\arcsec\ to WNW.\\
\noindent 280134: foreground gal.

\normalsize
\noindent {\bf A2295:}\\
\small
\noindent 270375: rising ORC; note low i; small companion gal 10\arcsec\ to E.; 
unfit for TF use; 5 min. integ.\\
\noindent 280131: background gal.\\
\noindent 280136: background gal.

\normalsize
\noindent {\bf A3744:}\\
\small
\noindent 610196: rising ORC; large extrap.\\
\noindent 610198: rising ORC.\\
\noindent 34131: faint disk; large extrapolation made to ORC.\\
\noindent 610220: heavy \hal\ absorption in one half of disk.\\
\noindent 610222: slightly rising ORC.\\
\noindent 34189: foreground gal.; patchy \HII\ distribution.\\
\noindent 610251: foreground gal.

\normalsize
\noindent {\bf A2457:}\\
\small
\noindent 320572: mostly bulge ORC; large correction applied to $W_{\rm ORC}$;
\NII\ line used.\\
\noindent 320573: outer isophotes differ in ell., PA w.r.t. inner isophotes.\\
\noindent 320574: background gal.\\
\noindent 320576: star 55\arcsec\ to NE; uncert. ellipticity; \NII\ patch for
radii $\lesssim$ 1.5\arcsec.\\
\noindent 320579: slightly rising ORC; uncert. ellipticity; \NII\ patch for 
2\arcsec\ $\lesssim$ radii $\lesssim$ 4\arcsec.\\
\noindent 320581: \NII\ patch for radii $\lesssim$ 2\arcsec.\\
\noindent 320583: asymm; merger?\\
\noindent 320591: foreground gal.; 5 min. integ.
\normalsize

In Fig. 7, we plot the ORCs folded about a kinematic center as described in 
Sec. 2.2.  The horizontal dashed line in each panel indicates the adopted (and 
uncorrected for inclination) half velocity width, $W$/2, for each galaxy and the 
vertical dashed line is drawn at \Ropt.  We have overlayed fits to the ORCs for the
cases in which shape corrections to the velocity widths are necessary.  Finally, in
Fig. 8 we give, as an 
example, the surface brightness profiles for the first 16 galaxies in Fig. 7.  
Again the vertical dashed line refers to \Ropt.  The solid line drawn along 
the disk is the fit to the disk over the range of radii assumed to cover the 
exponential portion of the disk.  The remainder of the plots of surface 
brightness profiles for the complete sample can be obtained by contacting the
first author.

Fig. 9 gives the ``raw'' TF plots of each cluster uncorrected for any cluster
incompleteness bias.  A computation of such bias will be presented in future 
work when data from all clusters is in hand.  Furthermore, the cluster systemic
redshifts used in obtaining these plots are preliminary.  Included in the TF plots 
is the template relation obtained from nearby clusters in G97b:
\be
y = -7.68x - 21.01
\ee
where $y$ is $M_{\rm cor}$ -- 5log$h$ and $x$ is log$W_{\rm cor}$ -- 2.5.

\acknowledgements
We thank Katie Jore for the use of her ORC fitting programs.  The results 
presented here are based on observations carried out at the Palomar Observatory 
(PO), at the Kitt Peak National Observatory (KPNO), at the Cerro Tololo 
Inter--American  Observatory (CTIO), the (late) 300' telescope of the National 
Radio Astronomy Observatory (NRAO), and the Arecibo Observatory, which is part of the National Astronomy and Ionosphere Center (NAIC).  KPNO AND CTI0 are 
operated by Associated Universities for Research in Astronomy and NAIC is operated 
by Cornell University, all under cooperative agreements with the National Science 
Foundation.  The Hale telescope at the PO is operated by the California Institute 
of Technology under a cooperative agreement with Cornell University and the Jet 
Propulsion Laboratory.  The National Radio Astronomy Observatory (NRAO) is
operated by Associated Universities, Inc. under a management agreement with NSF.
This research was supported by NSF grants AST94-20505 to RG and AST95-28960 to MH.  LEC was partially supported by FONDECYT grant \#1970735.

\newpage

\figcaption[Dale.fig1.ps] {The effect of a \NII\ data patch on a kinematically
folded \hal\ rotation curve lacking data in the central regions due to absorption.
The upper panel displays the \hal\ rotation curve for galaxy 400641 in Abell 119 without the patch, the lower panel includes the \NII\ data patch.  Notice the 
kinematical recentering changes the definition of the ORC origin in the spatial 
direction as well as in the velocity direction. \label{fig1}}

\figcaption[Dale.fig2.ps]{Sky and velocity distribution of galaxies in the 
clusters Abell 2806 and Abell 114.  Circles represent cluster members with 
measured photometry and widths; if unfilled, widths are poorly determined.  
Asterisks identify foreground and background galaxies and dots give the location 
of galaxies with known redshift, but lacking accurate width and/or photometry.  
Large square boxes indicate outlines of imaged fields with the 0.9 m telescope.  
Vertical dashed lines in the lower panels indicate 1 \rA\ (and 2 \rA\ for A114).  
The upper panels contain circles of radius 1 \rA\ (and 2 \rA\ for A114). 
\label{fig2}}

\figcaption[Dale.fig3.ps]{Sky and velocity distribution of galaxies in the 
clusters Abell 119 and Abell 194.  Filled circles, unfilled circles, asterisks, 
dots, large squares and dashed lines and circles follow the same convention as in
Fig. 2. \label{fig3}}

\figcaption[Dale.fig4.ps]{Sky and velocity distribution of galaxies in the 
clusters Abell 3193 and Abell 3381.  Filled circles, unfilled circles, asterisks, 
dots, large squares and dashed lines and circles follow the same convention as in
Fig. 2. \label{fig4}}

\figcaption[Dale.fig5.ps]{Sky and velocity distribution of galaxies in the 
clusters Abell 2295, Abell 2295b, and Abell 3744.  For A2295 and A3744, filled 
circles, unfilled circles, asterisks, dots, large squares and dashed lines and 
circles follow the same convention as in Fig. 2. For A2295b, filled triangles identify cluster members and 1 \rA\ is indicated by the dotted line in the lower panel and by the dotted circle of radius 1 \rA\ in the upper panel. \label{fig5}}

\figcaption[Dale.fig6.ps]{Sky and velocity distribution of galaxies in the cluster 
Abell 2457.  Filled circles, unfilled circles, asterisks, dots, large squares and 
dashed lines and circles follow the same convention as in Fig. 2. \label{fig6}}

\figcaption[Dale.fig7.ps]{\hal\ rotation curves for \Nopt\ galaxies (except for 
galaxies 400600 and 320572, for which the ORCs are obtained from a \NII\ emission 
line), folded about the kinematic centers.  The error bars include both the 
uncertainty in the wavelength calibration and the ORC fitting routine used.  Names 
of the galaxy and the corresponding parent cluster are given along with the CMB 
radial velocity.  Two dashed lines are drawn: the horizontal line indicates the 
adopted half velocity width, $W$/2, which in some cases arises from an 
extrapolation to the ORC or from a 21 cm width (see Table 2); the vertical line is at $R_{opt}$, the radius containing 83\% of the $I$ band flux.  For those cases
where a shape correction to the velocity width is used, a fit to the ORC is
indicated by a solid line.  Note that the rotation curves are {\it not} 
deprojected to an edge--on orientation. \label{fig7}}

\figcaption[Dale.fig8.ps]{A sampling of surface brightness profiles.  Names of the 
galaxy and the corresponding parent cluster are given in each panel.  Two lines are
drawn: the vertical dashed line is drawn at $R_{opt}$ and the solid line is an 
exponential fit to the disk, over the range of radii over which the disk is assumed
to behave exponentially. \label{fig8}}

\figcaption[Dale.fig9.ps]{``Raw'' TF plots for the 10 clusters are given.  We 
emphasize that the data have {\it not} been corrected for incompleteness bias.  
In the A2295 panel, the error bars containing filled circles represent members of
``A2295b,'' a cluster at a lower redshift than A2295.  The dashed line is the 
template relation valid for low $z$ clusters, Eqn. (4). \label{fig9}}

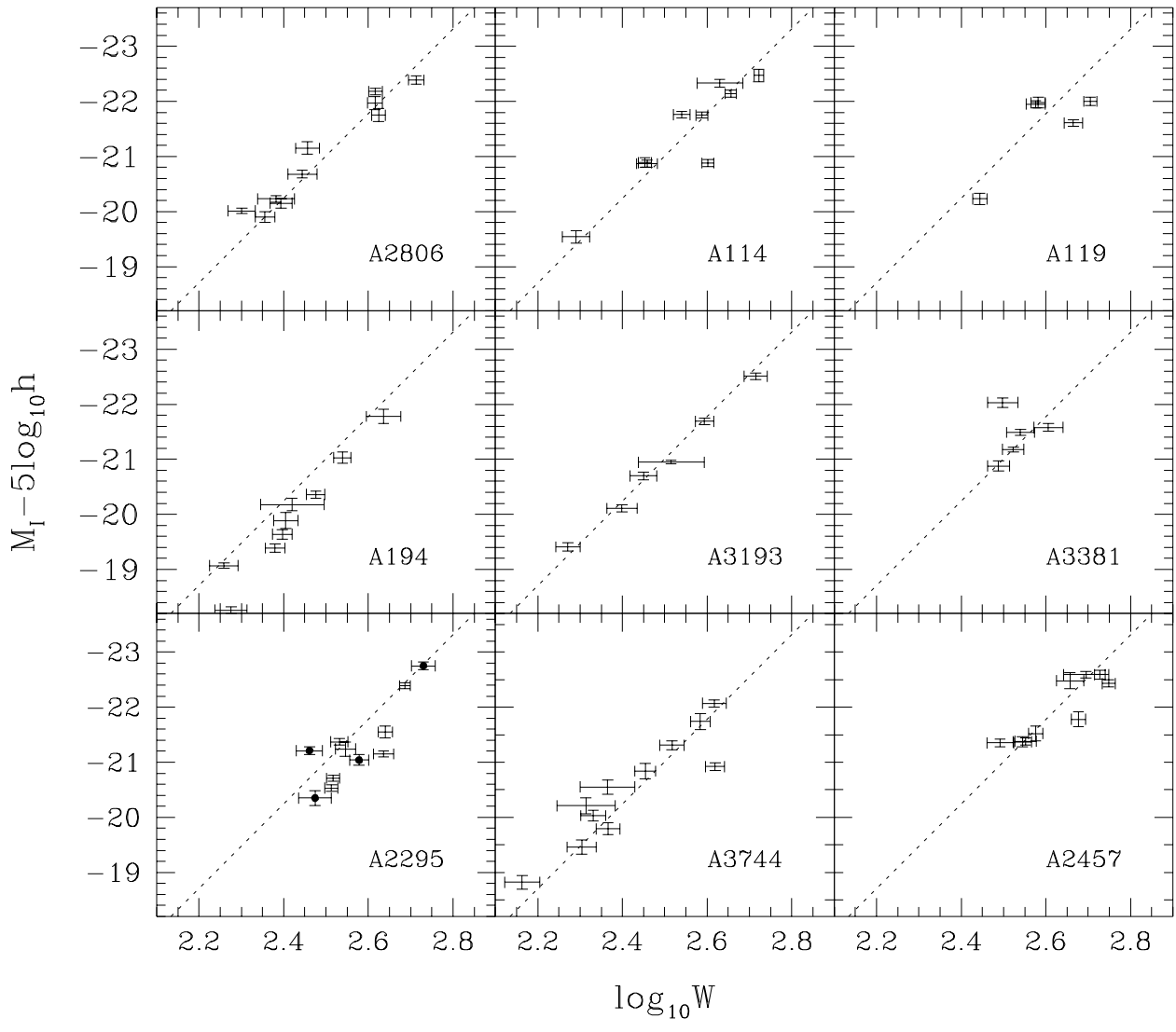
\end{document}